\documentclass[useAMS,usenatbib]{biom}
\usepackage[utf8]{inputenc}

\def\bSig\mathbf{\Sigma}

\newcommand{\bg}{\boldsymbol{g}}

\usepackage[figuresright]{rotating}
\usepackage{graphicx}
\usepackage{amsmath}   
\usepackage{amsfonts}
\usepackage{amssymb}   
\usepackage{bm}        
\usepackage{color}

\usepackage{setspace}

\title{Bayesian Non-Parametric Detection Heterogeneity in Ecological Models}

\author{D. Turek$^{*}$\email{dbt1@williams.edu} \\
	   Williams College, Williamstown, USA
	   \and 
	   C. Wehrhahn$^{*}$\email{cwehrhah@ucsc.edu}\\
	   University of California, Santa Cruz, USA
	   \and 
	   O. Gimenez$^{*}$\email{olivier.gimenez@cefe.cnrs.fr}\\
	   CNRS, France
	   }

\begin{document}

\begin{abstract}
Detection heterogeneity is inherent to ecological data, arising from factors such as varied terrain or weather conditions, inconsistent sampling effort, or heterogeneity of individuals themselves.  Incorporating additional covariates into a statistical model is one approach for addressing heterogeneity, but is no guarantee that any set of measurable covariates will adequately address the heterogeneity, and the presence of unmodelled heterogeneity has been shown to produce biases in the resulting inferences.  Other approaches for addressing heterogeneity include the use of random effects, or finite mixtures of homogeneous subgroups.  Here, we present a non-parametric approach for modelling detection heterogeneity for use in a Bayesian hierarchical framework.  We employ a Dirichlet process mixture which allows a flexible number of population subgroups without the need to pre-specify this number of subgroups as in a finite mixture.  We describe this non-parametric approach, then consider its use for modelling detection heterogeneity in two common ecological motifs: capture-recapture and occupancy modelling.  For each, we consider a homogeneous model, finite mixture models, and the non-parametric approach.  We compare these approaches using two simulation studies, and observe the non-parametric approach as the most reliable method for addressing varying degrees of heterogeneity.  We also present two real-data examples, and compare the inferences resulting from each modelling approach.  Analyses are carried out using the \texttt{nimble} package for \texttt{R}, which provides facilities for Bayesian non-parametric models.
\end{abstract}

%
%

\begin{keywords}
Bayesian non-parametrics; Capture-recapture; Detection heterogeneity; Markov chain Monte Carlo; \texttt{nimble}; \texttt{nimbleEcology}; Occupancy models; Statistical ecology.
\end{keywords}

\maketitle

\section{Introduction}

Inferring species distribution and demography are two key questions in ecology \citep{BegonEtAl2006}. However, addressing these questions is challenging when studying animals and plants in their natural environmental because of the inherent imperfection in the detection process of species or individuals.  When ignored, this issue of imperfect detection can lead to biased estimates of species distribution or survival \citep{gim2008,guru2017}. To cope with imperfect detection, occupancy models \citep{MackenzieEtAl2018} and capture-recapture models \citep{McCrea2014} were developed to provide unbiased estimates of species range and individual survival, with numerous applications in all fields of ecology.

As with any statistical model, inferences made from occupancy and capture-recapture analyses rely on assumptions which must be satisfied at least to a reasonable degree. Common to occupancy and capture-recapture models is the assumption of homogeneity of the detection process, which asserts that there is no unmodelled heterogeneity in species and individual detection probabilities.  If ignored, heterogeneity in detection leads to flawed inference in occupancy and capture-recapture models \citep{royle2006,gimenez_individual_2018}. Detection heterogeneity can be due to heterogeneous sampling effort, variation in animal abundance or behavior, site characteristics or even on account of varying observer skills. Ideally, covariates could be measured and incorporated in ecological models to account for detection heterogeneity. However, unexplained variation may still remain, or measuring the relevant covariates may simply be impossible.

When unmodelled heterogeneity exists, it can be accommodated using finite mixtures in which discrete latent variables are used to assign sites or individuals to mixture components (\emph{i.e.}, uniform groups) each characterized by group-specific parameters \citep{royle2006,pledger_open_2010,pledger_open_2003,louvrier_accounting_2018}. In simulation studies, finite mixtures were successful in decreasing bias in occupancy and survival probability estimates that was introduced by heterogeneity in the detection process \citep{pledger_performance,louvrier_accounting_2018}. However, from a practical perspective, the issue remains of selecting the number of mixture components for real-life data analyses, which is not straightforward \citep{cubaynes_assessing_2012,pohle_selecting_2017}.

Here, we propose a Bayesian non-parametric (BNP) approach to modeling heterogeneity in occupancy and capture-recapture models. BNP models provide a flexible approach that relaxes typical standard modeling assumptions, such as  the choice of a specific parametric kernel in density estimation, or here the choice of a fixed number of groups in finite mixtures. Previous uses of a BNP approach in ecological models include the modelling of wildlife migration patterns \citep{matechou_modelling_2017,diana_matechou}, the estimation of population size \citep{manrique,dorazio2008} and that of the probability of remaining in or vacating a given area \citep{ford2015modelling}. 

Perhaps the most widely used BNP model is the Dirichlet process mixture (DPM) model \citep{ferguson;73,ferguson;74,lo;84,escobar;94, escobar;west;95}, which is a mixture model with infinitely many components.  DPM models are a suitable fit for addressing the inherent heterogeneity present in ecological models.  We consider DPM models of Bernoulli distributions, with the distribution of the Bernoulli detection probability parameters arising from a Dirichlet process.  A DPM model can be represented in different, yet equivalent manners. Two of them are the the Chinese Restaurant Process \citep[CRP;][]{blackwell;mcqueen;73,Pi95,Pi96} and the  stick-breaking \citep[SB;][]{sethuraman;94} representations.

The flexibility of BNP models is usually translated into a hierarchical model, which relies on Markov chain Monte Carlo algorithms to sample from the resulting posterior distribution. The implementation of these algorithms is usually computationally complex and demanding in standard Bayesian software such as WinBUGS or JAGS \citep[e.g.][]{ohlssen_bnpwinbugs} and often requires writing specific code to implement the BNP model and sampling algorithms \citep{ford2015modelling}. Recently, the \texttt{nimble} \texttt{R} package introduced non-parametric functionality to address these difficulties, by supporting the use of non-parametric DPM models.  Specialized functions, distributions, and samplers are provided for both the CRP and SB representations. \citet{wehrhahn2018bayesian} contains additional details and examples of using these BNP modelling approaches in \texttt{nimble}.    

Here, we focus on the CRP representation to address detection heterogeneity in capture-recapture and occupancy models. The article is organized as follows. In the next section, we present homogeneous capture-recapture and occupancy models, finite mixtures and non-parametric models formulated in a Bayesian framework as hierarchical models. In Section 3, we give the MCMC sampling scheme used to fit non-parametric models and introduce the \texttt{nimble} and \texttt{nimbleEcology} \texttt{R} packages. Section 4 gives the results of two simulation studies, using capture-recapture and occupancy data, which validate the ability of our approach to capture detection heterogeneity in occupancy and survival probabilities. Section 5 illustrates our method using data from a study of gray wolf (\emph{canis lupus}), in trying to estimate occupancy and survival while accounting for detection heterogeneity. The final section provides general conclusions and discusses the potential of our approach.

\section{Models}
\label{sec:models}

We consider three different approaches to handling heterogeneity.  The first approach uses a homogeneous model, which disregards any heterogeneity and considers all individuals and sites to be identical.  The second approach uses finite mixtures of population subgroups, where individuals or sites within each subgroup are identical,  distinct population subgroups may differ in one or more characteristics, and the number of subgroups is pre-specified.  The homogeneous model is a special case of the finite mixture model, where the population contains only a single subgroup.

The third approach uses a non-parametric representation for modeling individual and species heterogeneity, where the number of population subgroups, the assignment of individuals and sites into subgroups, and the characteristics of each subgroup are determined by the data at the time of model fitting.  The non-parametric approach does not require pre-specifying a fixed number of population subgroups, but rather, the number of population subgroups itself is a model parameter.  We now describe the three model formulations in detail, and provide specifications of ecological capture-recapture and occupancy models using each.

\subsection{Homogeneous Models}
\label{sec:homogeneous_models}

In the homogeneous model, we assume that all $N$ observed individuals of the population are identical in their characteristics.  There is no heterogeneity between individuals or sites, and thus for any parameter  $\theta$ of the population, all individuals or sites share a common value of $\theta$.

\subsubsection*{Homogeneous Capture-Recapture Model} \ \\

We consider a basic ecological capture-recapture model, for binary-valued observation data of $N$ individuals or sites occurring over $T$ observational time periods.  We condition on the first observation of each individual occurring in time period $t=1$, although it is straightforward to relax this assumption to allow first-detections to occur in other time periods.  We parameterize the model in terms of survival probability between time periods $\phi$, and probability of detection conditional on being alive $p$.

In subsequent models we introduce heterogeneity in $p$, but in the homogeneous model all individuals are characterized by the constant probability of detection $p$.  We use a state-space formulation of the model, where binary states $x_{i,t}$ give the alive/dead status of individual $i$ at time $t$, and using binary observation data $y_{i,t}$ for site $i$ at time $t$.  The homogeneous capture-recapture model is written:

\begin{singlespace}
\begin{equation*}
\begin{aligned}
\phi &\sim \text{Uniform}(0, 1) \\
p &\sim \text{Uniform}(0, 1) \\
i = 1, &\ldots, \; N: \\
x_{i,1} &= 1 \\
x_{i,t} &\sim \text{Bernoulli}(\phi \cdot x_{i,t-1}) \text{ for } t = 2, \ldots, T \\
y_{i,t} &\sim \text{Bernoulli}(p \cdot x_{i,t}) \text{ for } t = 1, \ldots, T \\
\end{aligned}
\label{eq:model_cr_homogeneous_latent}
\end{equation*}
\end{singlespace}

\subsubsection*{Homogeneous Occupancy Model} \ \\

We consider a homogeneous static occupancy model for a total of $N$ sites, each observed at $T$ distinct sampling occasions.  We parameterize the model in terms of the constant site occupancy probability $\psi$, and probability of detection conditional on site occupancy $p$.  In subsequent models we introduce heterogeneity in $p$, but in the homogeneous model all sites share the probability of detection $p$.  We use a state-space formulation of the model, where binary states $z_{i}$ give the true occupancy status of site $i$, and using binary observation data $y_{i,t}$ for the observation of site $i$ at sampling occasion $t$.  The homogeneous occupancy model is written:

\begin{singlespace} \begin{equation*} \begin{aligned}
\psi &\sim \text{Uniform}(0, 1) \\
p &\sim \text{Uniform}(0, 1) \\
i = 1, &\ldots, \; N: \\
z_i &\sim \text{Bernoulli}(\psi) \\
y_{i,t} &\sim \text{Bernoulli}(p \cdot z_i) \text{ for } t = 1, \ldots, T \\
\end{aligned} \label{eq:model_occ_homogeneous_latent} \end{equation*} \end{singlespace}

\subsection{Finite Mixture Models}
\label{sec:finite_mixture_models}

In a finite mixture model specified as having $K \geq 2$ distinct population subgroups, each individual or site is considered to be a member of any of the $K$ subgroups with equal probability $1/K$.  We introduce a discrete indicator variable for each individual or site, $g_i$ for individual or site $i$, where $g_i$ denotes the ``group'' of individual or site $i$.  We use independent discrete uniform prior distributions over the set $\{1, 2, \ldots, K\}$ for each $g_i$, and $g_i = k$ indicates that individual or site $i$ is a member of population subgroup $k$.

Furthermore, each of the $K$ distinct subgroups may differ in one or more demographic characteristics.  Again considering the demographic parameter $\theta$, we introduce $K$ model parameters $\theta_1, \theta_2, \ldots, \theta_K$, which are given independent and identical prior distributions.  Then, all individuals or sites belonging to population subgroup $k$ display $\theta = \theta_k$.  To avoid issues of ``label exchanging'' between groups, and hence a lack of model identifiability, we also impose the constraint that $\theta_k \leq \theta_{k+1}$, or that the ordered set of parameters $\{\theta_1, \theta_2, \ldots, \theta_K\}$ is non-decreasing.

\subsubsection*{Finite Mixture Capture-Recapture Model} \ \\

We generalize the homogeneous capture-recapture model given in section \ref{sec:homogeneous_models} to a $K$-group finite mixture model, to allow heterogeneity in the probability of detection.  Specifically, we introduce $K$ new model parameters $p_1, p_2, \ldots, p_K$, where $p_k$ represents the probability of detection for individuals in subgroup $k$.  Each $p_k$ is given an independent $\text{Uniform}(0, 1)$ distribution, and we impose the constraint on these parameters that $p_{k} \leq p_{k^{\prime}}$ for $k < k^{\prime}$ to maintain identifiability.

Since $g_i$ gives the subgroup number which contains individual $i$, we use probability of detection $p_{g_i}$ for individual $i$.  Putting this together, the $K$-group finite mixture capture-recapture model is written as:

\begin{singlespace} \begin{equation*} \begin{aligned}
\phi &\sim \text{Uniform}(0, 1) \\
p_{k} &\sim \text{Uniform}(0, 1) \text{ for } k = 1, \ldots, K \\
p_{k} &\leq p_{k^{\prime}} \text{ for } k < k^{\prime} \\
i = 1, &\ldots, \; N: \\
g_i &\sim \text{DiscreteUniform}(\{1, \ldots, K\}) \\
x_{i,1} &= 1 \\
x_{i,t} &\sim \text{Bernoulli}(\phi \cdot x_{i,t-1}) \text{ for } t = 2, \ldots, T \\
y_{i,t} &\sim \text{Bernoulli}(p_{g_i} \cdot x_{i,t}) \text{ for } t = 1, \ldots, T \\
\end{aligned} \label{eq:model_cr_mixture_latent} \end{equation*} \end{singlespace}

\subsubsection*{Finite Mixture Occupancy Model} \ \\

We similarly generalize the homogeneous occupancy model from section \ref{sec:homogeneous_models} to a $K$-group finite mixture model using parameters $p_1, p_2, \ldots, p_K$, where $p_k$ represents the probability of detection for sites in subgroup $k$.  We use the same independent $\text{Uniform}(0, 1)$ prior distribution for each $p_k$, and impose the same constraint to ensure model identifiability.  Thus, $g_i$ indicates the subgroup which contains site $i$, which therefore has probability of detection $p_{g_i}$.  The $K$-group finite mixture occupancy model is written as:

\begin{singlespace} \begin{equation*} \begin{aligned}
\psi &\sim \text{Uniform}(0, 1) \\
p_{k} &\sim \text{Uniform}(0, 1) \text{ for } k = 1, \ldots, K \\
p_{k} &\leq p_{k^{\prime}} \text{ for } k < k^{\prime} \\
i = 1, &\ldots, \; N: \\
g_i &\sim \text{DiscreteUniform}(\{1, \ldots, K\}) \\
z_i &\sim \text{Bernoulli}(\psi) \\
y_{i,t} &\sim \text{Bernoulli}(p_{g_i} \cdot z_i) \text{ for } t = 1, \ldots, T \\
\end{aligned} \label{eq:model_occ_mixture_latent} \end{equation*} \end{singlespace}

\subsection{Non-Parametric Models}

Using a non-parametric approach, the pre-specification of a fixed number of subgroups is no longer required.  Furthermore, our previous assumption that individual or site assignments to subgroups must follow a discrete uniform distribution is also relaxed. In a non-parametric model the number of subgroups in the population is considered unknown and is inferred from the data. Theoretically, there could be an infinite number of population subgroups, although in practice there will never exceed $N$ subgroups.  This means there cannot exist more population subgroups than the total number of observations. But so as long as there are fewer than $N$ subgroups, then individuals or sites can probabilistically move out of an existing group and into to a newly created group, with its own distinct probability of detection.

As with the finite mixture model, the subgroup assignment structure is encoded in indicator variables $g_i$, where $g_i = k$ indicates that individual $i$ belongs to population subgroup $k$.  Consider the following conditional distribution for $g_i$, where $\bg_{1:(i-1)}=(g_1, \ldots, g_{i-1})$:

\begin{align} \label{eq:crpcond}
g_i \mid \bg_{1:(i-1)}, \alpha &\sim \left(\tfrac{1}{i-1+\alpha}\right) \sum_{j=1}^{i-1} \delta_{g_j} + \left(\tfrac{\alpha}{i-1+\alpha}\right) \, \delta_{g^{new}},
\end{align}



\noindent where $g^{new}$ is an integer not in $\bg_{1:i}$, $\alpha>0$ is the concentration parameter, and $\delta_x$ is a discrete measure concentrated at $x$, which translates to a point mass of probability located at $x$.  The discrete distribution (\ref{eq:crpcond}) for the group assignment of individual or site $i$ (conditional on the group assignments of individuals or sites $1, \ldots, i-1$) implies that each successive individual or site is assigned to an existing subgroup with probability proportional to the size of each subgroup, and is assigned to a new subgroup with probability proportional to $\alpha$.  The product of the successive conditional distributions given by~(\ref{eq:crpcond}) gives rise to the joint distribution of  $\bg_{1:N}$, which is the Chinese restaurant process (CRP) prior distribution with concentration parameter $\alpha$ \citep{blackwell;mcqueen;73,Pi95,Pi96}.  See \cite{li2019tutorial} for more details and interpretations of the CRP prior distribution. 

\begin{figure}
\centering
\includegraphics[width=3.4in]{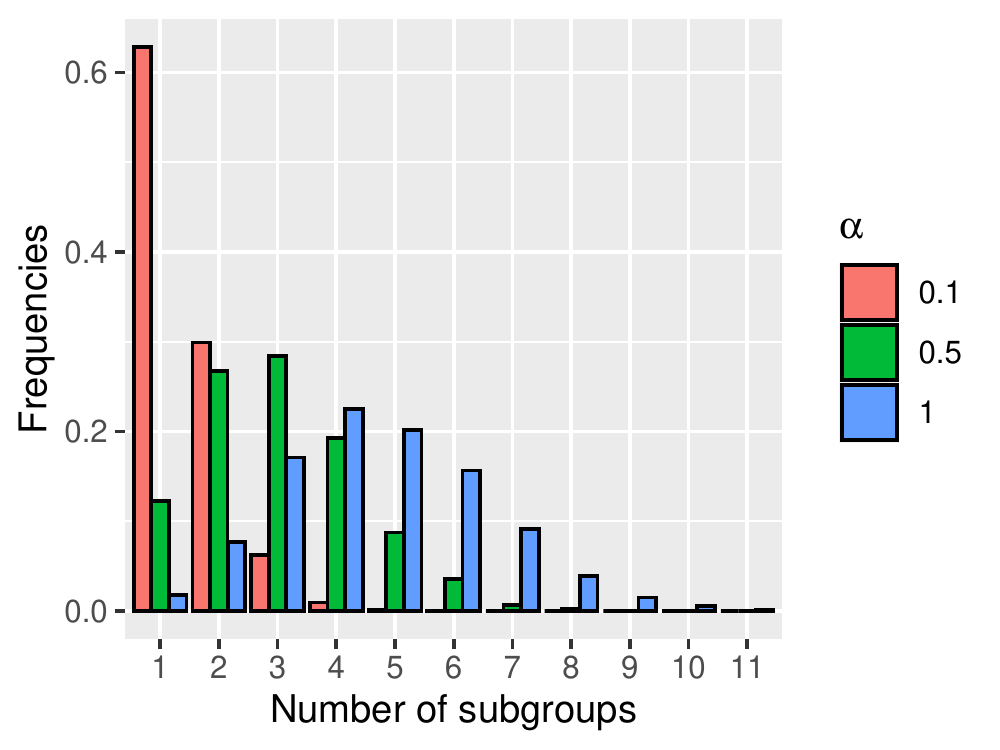}
\caption{Prior probability of number of subgroups induced by the CRP($\alpha$) prior distribution for different values of $\alpha$, when $N=60$.  A color version is available in the electronic version of this article.}
\label{fig:crpK}
\end{figure}

The strictly positive concentration parameter $\alpha$ of the CRP distribution influences the number of subgroups, through its control over the probability that individuals or sites are assigned into new subgroups.  The larger the value of $\alpha$, the more likely new subgroups are to be created.  Figure~\ref{fig:crpK} illustrates the effect of $\alpha$ on the number of subgroups created from the CRP prior distribution, when $N=60$. For $\alpha = 0.1$ generally only one or two subgroups are created, and infrequently three or more.  When $\alpha = 0.5$, the number of subgroups created generally falls between one and six.  For $\alpha = 1$, we seldom observe only one subgroup, and instead generally have between two and ten groups.  In our non-parametric models, rather than fixing $\alpha$ to a specific value, we will use a hyperprior distribution for $\alpha$ to allow the degree of heterogeneity within the dataset itself to dictate the plausible range for $\alpha$.

The possibility of the CRP prior using as many population subgroups as the total number of observations requires the inclusion of $N$ distinct demographic parameters $\theta_1, \ldots, \theta_N$ in our hierarchical non-parametric models.  These parameters are assumed to be independent and identically distributed, just as in the finite mixture models.



\subsubsection*{Non-Parametric Capture-Recapture Model} \ \\

In the non-parametric capture-recapture model, detection heterogeneity is flexibly addressed using a CRP prior distribution to allow for an unknown number of population subgroups. The number of population subgroups will thereby be inferred from the data, and is not fixed to a pre-specified value as in the finite mixture models described in section~\ref{sec:finite_mixture_models}.  We assign a $\text{CRP}(\alpha)$ prior distribution to $\bg_{1:N}$, the vector of subgroup indicator variables.  We use a $\text{Gamma}(1,1)$ hyperprior distribution for the CRP concentration parameter $\alpha$.

We introduce probabilities of detection $p_1,  p_2, \ldots, p_N$, each independent and identically following a $\text{Uniform}(0,1)$ prior distribution. In general, fewer than $N$ subgroups are actually used by the CRP prior distribution, and thus only a subset of the probabilities of detection are ``active'' in terms of their influence on the model likelihood calculation.  As in the finite mixture models, $g_i = k$ indicates that individual $i$ is a member of subgroup $k$, and therefore has probability of detection $p_{g_i}$.  The full non-parametric capture-recapture model is written as:

\begin{singlespace} \begin{equation*} \begin{aligned}
\phi &\sim \text{Uniform}(0, 1) \\
p_{k} &\sim \text{Uniform}(0, 1) \text{ for } k = 1, \ldots, N \\ 
\alpha &\sim \text{Gamma}(1, 1) \\
\bm{g}_{1:N} &\sim \text{CRP}(\alpha) \\
i = 1, &\ldots, \; N: \\
x_{i,1} &= 1 \\
x_{i,t} &\sim \text{Bernoulli}(\phi \cdot x_{i,t-1}) \text{ for } t = 2, \ldots, T \\
y_{i,t} &\sim \text{Bernoulli}(p_{g_i} \cdot x_{i,t}) \text{ for } t = 1, \ldots, T \\
\end{aligned} \label{eq:model_cr_nonparametric_latent} \end{equation*} \end{singlespace}

\subsubsection*{Non-Parametric Occupancy Model} \ \\

Similarly, we generalize the finite mixture occupancy model to use a $\text{CRP}(\alpha)$ prior distribution for individual group assignments $\bg_{1:N}$, and assign a $\text{Gamma}(1,1)$ hyperprior distribution for $\alpha$.  We include the maximum possible necessary number of distinct probabilities of detection $p_1, p_2, \ldots, p_N$, each with independent $\text{Uniform}(0,1)$ prior distributions.  Here, $g_i = k$ indicates that site $i$ is a member of subgroup $k$, and therefore has probability of detection $p_{g_i}$.  The full non-parametric occupancy model is written as:

\begin{singlespace} \begin{equation*} \begin{aligned}
\psi &\sim \text{Uniform}(0, 1) \\
p_{k} &\sim \text{Uniform}(0, 1) \text{ for } k = 1, \ldots, N \\ 
\alpha &\sim \text{Gamma}(1, 1) \\
\bm{g}_{1:N} &\sim \text{CRP}(\alpha) \\
i = 1, &\ldots, \; N: \\
z_i &\sim \text{Bernoulli}(\psi) \\
y_{i,t} &\sim \text{Bernoulli}(p_{g_i} \cdot z_i) \text{ for } t = 1, \ldots, T \\
\end{aligned} \label{eq:model_occ_nonparametric_latent} \end{equation*} \end{singlespace}

\section{Model Fitting via Markov chain Monte Carlo}
\label{sec:model_fitting_mcmc}

Mathematical descriptions of our approaches to modeling heterogeneity within a population were given in section \ref{sec:models}.  These models are formulated in a Bayesian framework as hierarchical models.  Prior distributions are specified for model parameters, and the conditional distributions for discrete latent states (\emph{e.g.,} dead/alive status) and the data likelihoods are given.  The general tool for fitting hierarchical models to data is Markov chain Monte Carlo \citep[MCMC;][]{brooks_handbook_2011}, a stochastic sampling algorithm for generating posterior samples for model parameters, conditional on the data.  Demographic inferences are subsequently performed using the empirical posterior distributions of model parameters.

Markov chain Monte Carlo is a powerful tool for fitting hierarchical models to data, but the inherent mathematical complexity of MCMC sampling often requires the use of software.  We make use of the recently developed \texttt{nimble} \texttt{R} package \citep{nimble-software:2019}, which offers new degrees of freedom for algorithmic development and customization of MCMC sampling \citep{de2017programming}. The ability to write custom distributions for use in hierarchical models, and the flexible nature of \texttt{nimble}'s MCMC engine has provided noteworthy gains in fitting ecological models \citep{turek2016efficient}, and more generally in the study of MCMC algorithms \citep{turek2017automated}.

We make use of \texttt{nimble}'s MCMC engine for fitting these hierarchical models, after expressing them in the BUGS language \citep{lunn_bugs_2009}.  In addition, we make use of two aspects of \texttt{nimble}'s flexibility.  For modeling of individual heterogeneity, we make use of the BNP distributions and corresponding sampling algorithms, which are a recent addition to the \texttt{nimble} package.  Second, we enhance performance of the specific ecological models by using custom likelihood distributions provided by the \texttt{nimbleEcology} package \citep{nimble-ecology-software:2020} to remove latent states from the model structures.

\subsection{Non-Parametric Distributions and MCMC Sampling}

When a hierarchical model is formulated using the CRP distribution, the \texttt{dCRP} distribution (available in the \texttt{nimble} package) assigns the joint prior distribution arising from (\ref{eq:crpcond}) to the labeling vector, $\bg$. Correspondingly, a specialized sampler is assigned by \texttt{nimble}'s MCMC engine. Because the likelihood function of the ecological models presented in section~\ref{sec:models} are not conjugate for the prior distribution  of the cluster parameters, $p_k$, the non-conjugate sampling algorithm described in Algorithm 8 of~\cite{Ne00} is assigned to $\bg$.


As described in section~\ref{sec:models}, under a non-parametric approach, the number of population subgroups is not fixed. In terms of the MCMC sampling scheme this means that the number of subgroups, and therefore the number of cluster parameters which are active, can vary with every MCMC iteration. As \texttt{nimble} does not support dynamic length allocation, the number of cluster parameters defined in the model must be fixed. A safe option would be to consider $N$ cluster parameters, however this is highly inefficient both in terms of computation and storage, especially for large values of $N$. To reduce this inefficiency \texttt{nimble} allows the specification of $N^{\prime}<N$ cluster parameters. If upon any MCMC iteration more than $N^{\prime}$ groups are created, then a warning is issued. Additionally, to reduce the computational burden of the non-parametric sampling, only the active cluster parameters, $p_k$, are updated.       

As  discussed in section~\ref{sec:models}, the concentration parameter $\alpha$ has important implications in the clustering structure of the model. Therefore, efficient sampling of $\alpha$ is an important matter. Although its posterior distribution does not belong to any known class of distributions, when a gamma prior distribution is considered for $\alpha$, a computationally efficient sampling scheme \citep[][section 6]{escobar;west;95} is assigned by \texttt{nimble}'s MCMC engine.

\subsection{Likelihood Distributions Using \texttt{nimbleEcology}}

To reduce computation time, we make use of the \texttt{nimbleEcology} \texttt{R} package \citep{nimble-ecology-software:2020} for specifying the ecological hierarchical models.  The \texttt{nimbleEcology} package provides likelihood distributions specific for a variety of common ecological models, which are implemented as custom distributions using the \texttt{nimble} package.  The likelihood distributions provided in \texttt{nimbleEcology} include those for capture-recapture models, occupancy and dynamic occupancy models, and more generally for discrete hidden Markov models (HMMs) as appear in multi-state or multi-event capture-recapture \citep[\emph{e.g.,}][]{turek2016efficient}.

For each type of ecological model, the likelihood distribution provided by \texttt{nimbleEcology} marginalizes over discrete latent states to directly calculate the unconditional likelihood of observed data.  This allows removal of discrete latent state variables -- $x_{i,t}$ alive/dead indicator variables in capture-recapture, and $z_{i}$ occupancy indicator variables in occupancy modelling -- from the hierarchical model.  This reduces model size and the necessary model computations, and increases the speed of MCMC mixing of top-level model parameters ($\phi$ in capture-recapture, and $\psi$ in occupancy models) to generate stronger posterior inferences in less computational time.


Using the distributions provided in \texttt{nimbleEcology} to remove latent states does not alter the posterior results generated from each model; it only serves to increase the speed of generating inferences.  The only noteworthy difference is that posterior inference for the discrete latent states \emph{cannot} be performed, since samples for these latent states are never generated.  So, for example, we could not perform inference for the alive/dead status $x_{i,t}$ of specific individuals in the capture-recapture setting.

\section{Simulations}

We undertake two simulation studies to assess performance of the various approaches to modeling individual and site heterogeneity.  The first study is in the context of capture-recapture models, and the second study in that of occupancy models.

Both simulations consider the effect of varying degrees of heterogeneity in detection probability between two population subgroups on the accuracy of inferences. In each, we fit the homogeneous model, 2-group and 3-group finite mixture models, and the non-parametric model as were presented in section \ref{sec:models}.  \texttt{R} code for all simulations, including the \texttt{nimble} specifications for each model using the likelihood distributions provided in the \texttt{nimbleEcology} package, are provided as supplemental material.

Next we describe the details of each simulation, and then present results.

\subsection{Capture-Recapture Simulation}


In the capture-recapture simulation, we consider two population subgroups each with 800 individuals, for a total population size of 1,600.  We use eight observational periods, and condition on all individuals being sighted on the first observational period.  Survival probability is fixed at $\phi = 0.7$ for simulating data, and we focus on the ability of various modeling approaches to estimate survival.

One population subgroup is fixed as having individual probability of detection $p_0 = 0.8$, where detection is conditional on being alive.  Individuals in the other subgroup have a fixed detection probability $p$, which varies between simulations.  We consider values of $p$ between $0.1$ and $0.8$ in increments of $0.1$, where the terminal case $p=0.8$ coincides with the detection probability of the first subgroup, and therefore represents a homogeneous population.

\subsection{Occupancy Simulation}


In the occupancy simulation, we consider two subgroups each with 2,000 sites, for a total of 4,000 sites.  We use six independent observations of each site.  The proportion of occupied sites is fixed at $\psi = 0.7$ for simulating data, and we focus on the ability of various modeling approaches to estimate the true occupancy proportion.

One subgroup is fixed as having probability of detection $p_0 = 0.8$ on each independent site visit, where detection is conditional on a site being occupied.  The other subgroup has a fixed detection probability $p$, which varies between simulations.  We consider values of $p$ between $0.1$ and $0.8$ in increments of $0.1$, but with a finer resolution on the interval between $0.1$ and $0.4$.  Again, the terminal case $p=0.8$ coincides with the detection probability of the first subgroup, and therefore represents a homogeneous population.

\subsection{Simulation Results}

Posterior inferences were performed for individual survival probability $\phi$ in the capture-recapture simulation and for site occupancy proportion $\psi$ in the occupancy simulation.  We use the posterior median and equal-tailed 95\% Bayesian credible intervals for inferences under each model.  Fitted models include a homogeneous model (Hom), 2-group and 3-group finite mixture models (FM 2 and FM 3, respectively), and a non-parametric model (NP).  Each model was fit to simulated data using MCMC, as described in section \ref{sec:model_fitting_mcmc}.

\begin{figure}[ht!]
\centering
\includegraphics[width=3.4in]{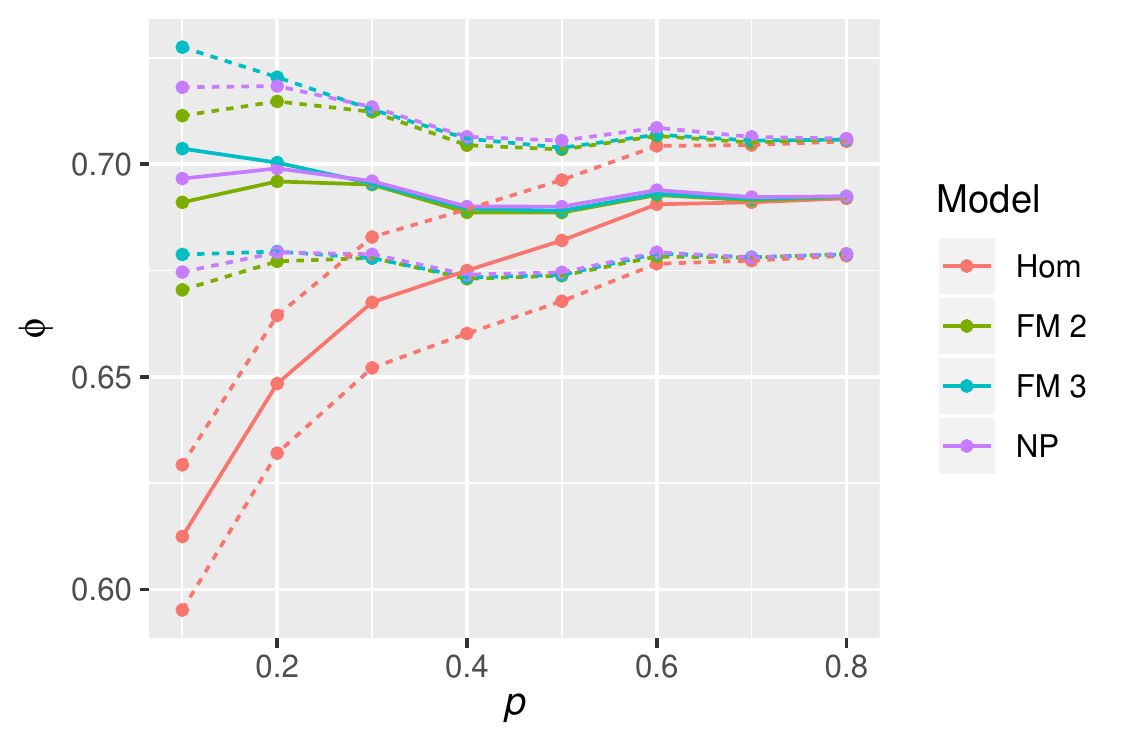}
\caption{Capture-recapture simulation results for survival probability $\phi$ using different models for heterogeneity in detection probability.  $N = 1,600$ individuals are simulated between two population subgroups, one with probability of detection $p_0 = 0.8$, and the other with probability of detection $p$.  True survival probability is fixed at $\phi = 0.7$.  Solid lines show posterior median estimates of $\phi$ from each model, and dashed lines show upper and lower limits of a 95\% Bayesian credible interval.  A color version is available in the electronic version of this article.}
\label{fig:simulation_cr}
\end{figure}

Results for the capture-recapture simulation appear in Figure \ref{fig:simulation_cr}.  The homogeneous model consistently under-estimates $\phi$, more severely for larger discrepancies in detection probability between the two groups, with the discrepancy diminishing and disappearing as the detection probabilities of the two groups converge.  For the lowest value of $p=0.1$, and hence a large difference in detection probability between groups, the 3-group mixture model has a slight tendency to inflate estimates of $\phi$, while the opposite is true of the 2-group mixture model; however this difference is minor and quickly disappears as the group difference deceases.  Otherwise, the mixture models and non-parametric models are all similarly successful in generating accurate inferences of $\phi$.  We also note that all models exhibit a slight negative bias in their estimates of $\phi$, which diminishes as $p_0$ and $p$ jointly approach one.


\begin{figure}[ht!]
\centering
\includegraphics[width=3.4in]{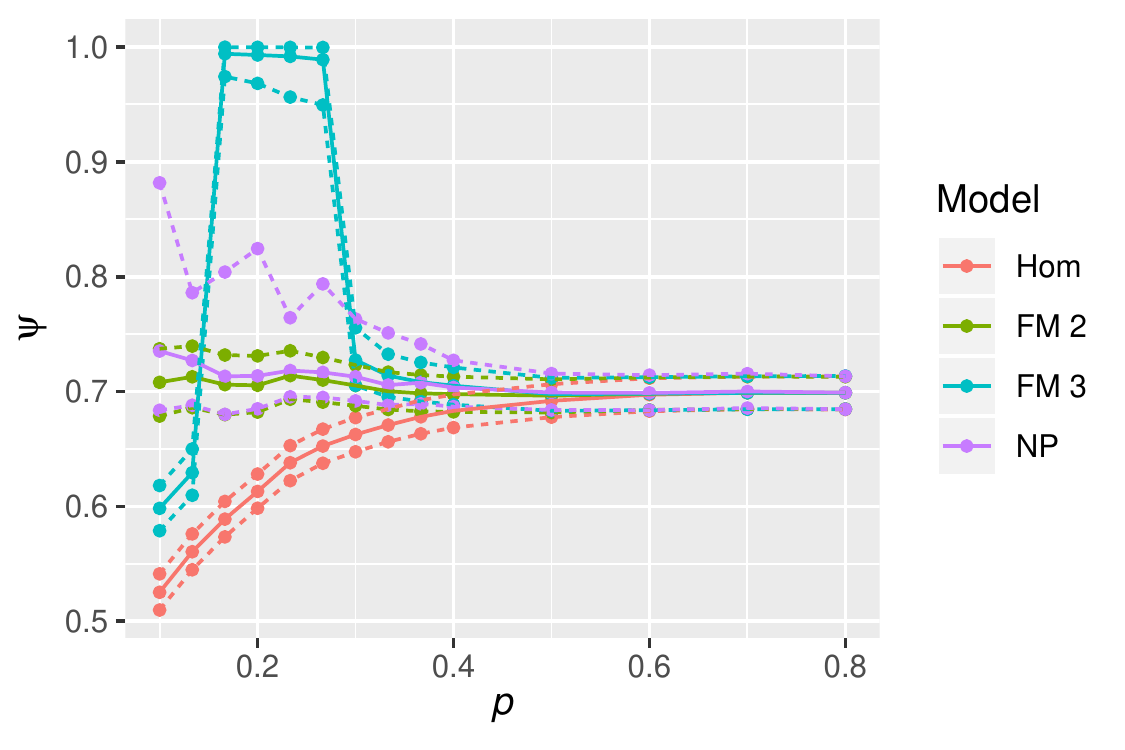}
\caption{Occupancy simulation results for occupancy $\psi$ using different models for heterogeneity in detection probability.  $N = 4,000$ individuals are simulated between two population subgroups, one with probability of detection $p_0 = 0.8$, and the other with probability of detection $p$.  True occupancy is fixed at $\psi = 0.7$.  Solid lines show posterior median estimates of $\psi$ from each model, and dashed lines show upper and lower limits of a 95\% Bayesian credible interval.  A color version is available in the electronic version of this article.}
\label{fig:simulation_occ}
\end{figure}

Results for the occupancy model simulation appear in Figure \ref{fig:simulation_occ}.  Once again, for low values of $p$ the homogeneous model underestimates $\psi$, more severely for larger discrepancies between the two groups.  We also see a regime of $p$ values between (approximately) 0.167 and 0.267, in which the 3-group mixture model vastly over-estimates $\psi$, with posterior median estimates in excess of 0.99.  The precise location of this regime varied somewhat depending on simulation parameters (specifically the number of individuals in the population, and number of observation periods) but its existence persisted in all simulations.  That is, there exists the potential for highly inaccurate inferences when the number of groups in a finite mixture model is chosen either too high, or too low.

The 2-group mixture model (which is also the data-generating model) generated reliable inferences for all values of $p$.  The non-parametric model admitted greater uncertainty for the upper limit of the 95\% credible interval, especially for lower values of $p$, but the posterior median estimates generated using the non-parametric model were consistently near to the true parameter value $\psi = 0.7$.

\section{Examples}

We present two real-data examples, one in the context of capture-recapture and the other in that of occupancy modeling.  For each example, we fit a homogeneous model for detection probability, finite mixture models containing between two and ten population subgroups, and a non-parametric model.  In addition to parameter inferences under each model, we also present the WAIC \citep{watanabe2010asymptotic, gelman2014understanding} of each fitted model.  The WAIC value is a measure of the goodness-of-fit of a hierarchical model, calculated using chains of posterior MCMC samples.  WAIC is measured on the scale of deviance, and therefore lower values of WAIC indicate a more parsimonious fit to the data.  Finally, for each example we also present the posterior distribution for the number of population subgroups in the non-parametric model.  This suggests at the degree of heterogeneity present in the data as inferred using the non-parametric model, the only model which does not pre-define the number of population subgroups.  The datasets used for each example are available on GitHub at: \texttt{https://github.com/danielturek/bnp-examples-data}.

\subsection{Capture-Recapture Example}


We consider wolf (\emph{canis lupus}) capture-recapture data collected in France between 1995 and 2003, as studied in \citet{cubaynes_importance_2010}.  The original data contains binary detection data for a total of 87 wolves, over $T = 8$ observation periods.  Since we condition on the first sighting of each individual, we excluded the twenty individuals who were first observed on the final observation period.  This leaves a total of $N = 67$ unique individuals in the dataset.  Models for detection heterogeneity described in section \ref{sec:models} were fit to this data using MCMC, and inference was performed for individual survival probability $\phi$ under each model.

\begin{table}[!ht]
\caption{Capture-recapture example results.  Posterior inferences are for individual survival probability $\phi$, and WAIC values indicate the goodness-of-fit of each model.}
\label{tab:example_cr}
\begin{center}
\begin{tabular}{lccc}
Model & Median & 95\% BCI & WAIC \\
\hline
Homogeneous & .80 & (.70, .89) & 237.5 \\
Finite Mixture 2 & .91 & (.79, .99) & 209.1 \\
Finite Mixture 3 & .91 & (.80, .99) & 201.9 \\
Finite Mixture 4 & .92 & (.80, .99) & 198.3 \\
Finite Mixture 5 & .92 & (.80, .99) & 198.3 \\
Finite Mixture 6 & .91 & (.80, .99) & 199.9 \\
Finite Mixture 7 & .91 & (.79, .99) & 200.6 \\
Finite Mixture 8 & .90 & (.79, .99) & 201.4 \\
Finite Mixture 9 & .90 & (.79, .98) & 202.1 \\
Finite Mixture 10 & .90 & (.78, .98) & 202.8 \\
Non-Parametric & .92 & (.81, .99) & 199.8 \\
\end{tabular}
\end{center}
\end{table}

\ \\

Posterior median and 95\% credible intervals, as well as the WAIC value of each fitted model are presented in Table \ref{tab:example_cr}.  The homogeneous model produces the largest WAIC value among those models considered (indicating the poorest fit to the data), and the lowest estimates of $\phi$.  Posterior inferences from all mixture models and the non-parametric model are nearly indistinguishable, with median posterior values for $\phi$ around 0.91, and 95\% credible intervals of approximately $(0.80, 0.99)$.  We note that the non-parametric model produces the third lowest WAIC value, being 1.5 higher than the lowest two WAIC values (equal from the 4-group and 5-group mixture models).  Further, inferences for $\phi$ generated under the non-parametric model are nearly identical to those of the 4-group and 5-group mixture models.  Given our uncertainty in the structure and degree of heterogeneity, the non-parametric model provides defensible inferences and goodness-of-fit.


\begin{figure}[ht!]
\centering
\includegraphics[width=3.4in]{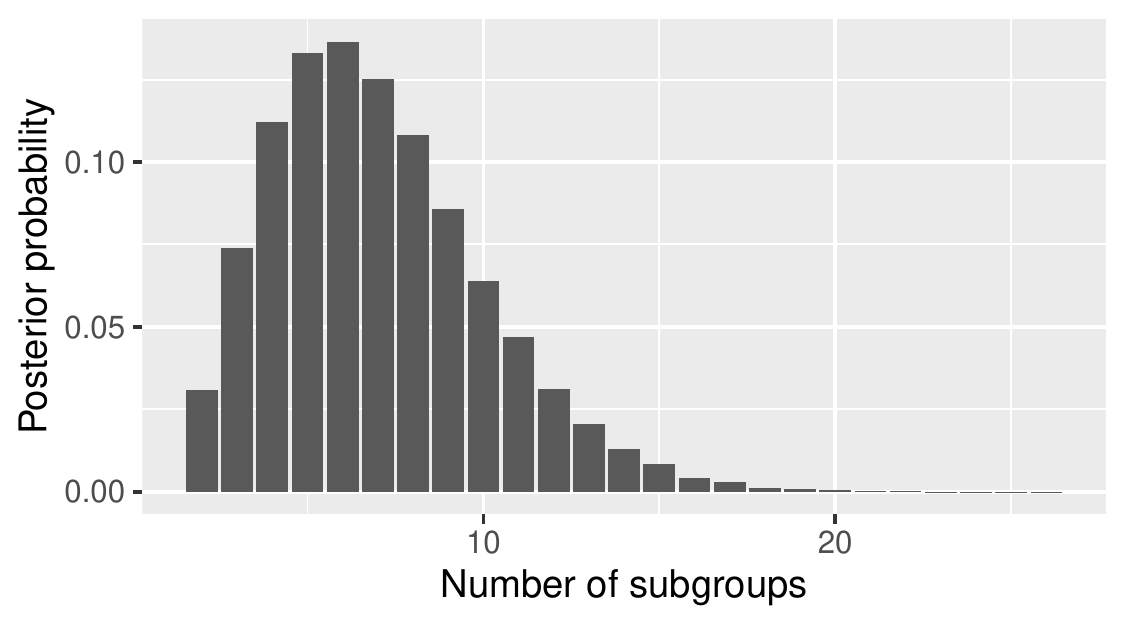}
\caption{Posterior distribution of the number of population subgroups under the non-parametric model, for the capture-recapture example.}
\label{fig:example_cr_bnp_groups}
\end{figure}

Figure \ref{fig:example_cr_bnp_groups} displays the posterior distribution for the number of population subgroups, as inferred by the non-parametric model.  We observe a right-skewed distribution placing the bulk of the posterior mass roughly between three and ten subgroups, with a posterior median of seven, the mode appearing at six subgroups, and a 90\% Bayesian credible interval for the number of subgroups as $(3, 13)$.  This distribution suggests that our consideration of finite mixture models containing between two and ten subgroups was a reasonable choice, even though this was an uninformed and somewhat arbitrary selection.

\subsection{Occupancy Example}


For our occupancy example, we consider a second wolf (\emph{canis lupus}) dataset \citep{louvrier_accounting_2018} collected in France in 2013.  Opportunistic observational data such as tracks, scat, and prey remains were collected from $N = $ 3,211 grid cells, each being a 10km $\times$ 10km square.  Each site was surveyed on a total of $T = 4$ independent observation occasions.  The categorical data, in total, consisted of 250 ``unambiguous detections'', 54 ``ambiguous detections'', and 12,540 ``non-detections''.  We convert this to binary data, wherein both ``unambiguous'' and ``ambiguous'' detections are considered to be positive detections.  Models for detection heterogeneity described in section \ref{sec:models} were fit to this data using MCMC, and inference was performed for site occupancy proportion $\psi$ under each model.

\begin{table}[!ht]
\caption{Occupancy model example results.  Posterior inferences are for site occupancy proportion $\psi$, and WAIC values indicate the goodness-of-fit of each model.}
\label{tab:example_occ}
\begin{center}
\begin{tabular}{lcccc}
Model & Median & 95\% BCI & WAIC \\
\hline
Homogeneous & .063 & (.054, .073) & 2237.0 \\
Finite Mixture 2 & .079 & (.063, .101) & 2195.8 \\
Finite Mixture 3 & .087 & (.067, .122) & 2178.3 \\
Finite Mixture 4 & .092 & (.068, .139) & 2168.7 \\
Finite Mixture 5 & .089 & (.068, .132) & 2172.4 \\
Finite Mixture 6 & .090 & (.069, .130) & 2172.1 \\
Finite Mixture 7 & .089 & (.068, .125) & 2173.7 \\
Finite Mixture 8 & .087 & (.068, .122) & 2174.9 \\
Finite Mixture 9 & .087 & (.068, .119) & 2175.9 \\
Finite Mixture 10 & .086 & (.069, .119) & 2175.4 \\
Non-Parametric & .090 & (.066, .359) & 2166.7 \\
\end{tabular}
\end{center}
\end{table}

\ \\

Posterior median and 95\% credible intervals, as well as the WAIC value of each fitted model are presented in Table \ref{tab:example_occ}.  The homogeneous model again produces the largest WAIC value (indicating the poorest fit to the data), and the lowest estimates of $\psi$.  The 2-group mixture model yields the second-highest WAIC value and also lower estimates of $\psi$ than the remaining models, and the 3-group mixture model yields the third highest WAIC value.  This suggests a nontrivial degree of heterogeneity in detection probability between sites.

The remaining mixture models ($K \geq 4$ groups) give WAIC values between 2168.7 and 2175.9, and exhibit small variations in the inferences for $\psi$, in particular in the upper limit of the 95\% credible interval.  In contrast, the non-parametric model yields the lowest WAIC value among all models (2166.7) indicating the best fit to the data.  The posterior median estimate from the non-parametric model is similar to that of the $K \geq 4$ mixture models, but the credible interval is wider.  In particular, the non-parametric model suggests a higher upper-bound for the 95\% credible interval for $\psi$ than any other model considered.  This result may be reasonable, since WAIC suggests the non-parametric model provides the best fit to the data, and we are completely uncertain as to the degrees of detection heterogeneity which exists in this opportunistic sampling dataset.


\begin{figure}[ht!]
\centering
\includegraphics[width=3.4in]{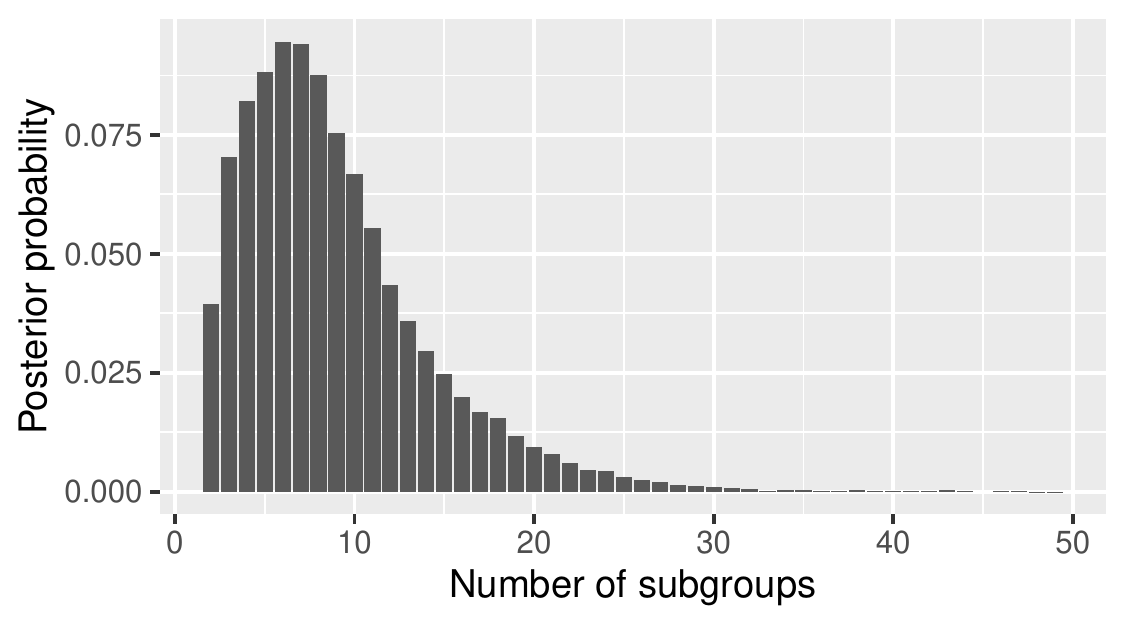}
\caption{Posterior distribution of the number of population subgroups under the non-parametric model, for the occupancy model example.}
\label{fig:example_occ_bnp_groups}
\end{figure}

We again present the posterior distribution for the number of population subgroups as inferred by the non-parametric model in Figure \ref{fig:example_occ_bnp_groups}.  This distribution is again right-skewed, this time spanning a wider range for plausible numbers of subgroups.  This makes intuitive sense, as our occupancy example data considers a much larger number of observations (3,211 sites, rather than a total of 67 individuals in the capture-recapture example).  The posterior distribution has a median of eight, once again the mode appears at six subgroups, and has a 90\% Bayesian credible interval of $(3, 19)$.

\section{Discussion}

Here, we have employed a Bayesian non-parametric approach to modeling heterogeneity in the detection process of ecological models.  Our approach used a Chinese restaurant process (CRP) prior distribution to model group memberships, which has the benefit of not requiring specification of the number of population sub-groups \emph{a priori}.  Using the CRP prior, both the distribution of individuals or sites among distinct subgroups and the number of subgroups are inferred from the data.  This strategy obviates any process of model selection used to choose a ``best value'' for the number of groups $K$ when using a finite mixture model.

The tendency of the CRP distribution towards using fewer (and hence more diverse) subgroups, or towards using a larger number of tightly specified subgroups is governed by a concentration parameter ($\alpha$).  The effect of different values of $\alpha$ is seen in Figure \ref{fig:crpK}, although we avoid the need to choose a particular value of $\alpha$ by specifying an uninformative Gamma prior distribution.  This approach of using a CRP prior distribution to model heterogeneity is used in diverse areas of research including topic modeling \citep{blei2010nested}, genomics \citep{qin2006clustering}, and evolutionary clustering \citep{ahmed2008dynamic}, among others.

The examples and simulations used herein focus on the most basic forms of capture-recapture and occupancy models.  Specifically, we have considered the Cormack-Jolly-Seber capture-recapture model, and a single-season static occupancy model.  Although our examples focused on the basic forms of each model, the same technique are readily applied to more complex variations of each model.  For example, we could apply the same non-parametric CRP prior distribution in multi-state capture-recapture, or in dynamic, multi-season, or multi-species occupancy models.  Similarly, one can readily extend the hierarchical models to incorporate individual or environmental covariates affecting population demographics, survey effort, or the detection processes.  The approach we have used is quite general, and widely applicable to ecological models where population heterogeneity is a consideration.

We have used the \texttt{nimble} \texttt{R} package to specify the hierarchical models described herein, and to fit these models to data using MCMC.  \texttt{nimble} provides degrees of flexibility which are not available in other software packages.  Specifically, we leverage \texttt{nimble}'s ability to use custom-written likelihood distributions in a hierarchical model, specifically the ecological likelihood distributions provided in the \texttt{nimbleEcology} package.  Further, \texttt{nimble} provides the ability to specify the sampling algorithms used by the MCMC, and even to oneself write customized sampling algorithms for use in the MCMC.  Indeed, the MCMC sampling algorithms used for fitting our models -- specifically those used for the CRP concentration parameter $\alpha$, and for the CRP-distributed group membership indicators $\bm{g}_{1:N}$ -- are themselves custom sampling algorithms written for precisely these non-parametric motifs, and added into \texttt{nimble}'s MCMC repertoire of algorithms.  That said, \texttt{nimble} does not attempt to provide ``canned'' algorithms, nor any particular pre-written model structures, but rather an environment for writing custom functions, statistical algorithms, and distributions, and the application of these to generally-specified hierarchical model structures.  The goal of \texttt{nimble} is to provide a flexible model and algorithmic programming environment to facilitate highly efficient analysis of models and complex data.


It is common that heterogeneity will be present to some degree in the detection process of ecological models.  In practice, this may be detected by goodness-of-fit tests \citep{jeyam_gof}, or perhaps based on prior expert knowledge. When detection heterogeneity is known or suspected to be present, and suitable covariates are not available to accurately model this heterogeneity, we recommend using a BNP modelling approach.  This approach alleviates the necessity of selecting the number of components used in a finite mixture model, which is an inherently difficult and oftentimes subjective process.  No less, the exact number of mixture components is generally not the primary inferential focus.  Use of a BNP modelling approach, as demonstrated herein, accounts for whatever degree of heterogeneity may be present while requiring no subjective choices or guesswork.  This provides an effective approach to reducing bias in the resulting demographic inferences.


\section*{Acknowledgements}

Support for DT was provided by Fulbright Research Scholarship Award 9183-FR and also the Williams College Class of 1945 World Fellowship.  Support for CW was partially provided by award NSF-DMS 1622444.  Support for OG was provided by a grant from CNRS and ``Mission pour l'interdisciplinarit\'{e}'' through its ``Osezl'interdisciplinarit\'{e} call.''  We warmly thank the French Office of Biodiversity (OFB) for sharing the wolf datasets.

\bibliographystyle{biom} 
\bibliography{biblio,biblio_DT}

\appendix
\newpage
\section{Simulation Code}
\label{app:code}

\begin{singlespace}
\begin{footnotesize}
\begin{verbatim}
library(nimble)
library(nimbleEcology)


##################################################
############ Capture-Recapture Models ############
##################################################

## number of individuals
N <- 800
## number of observation periods
T <- 8
## time period of first capture
first <- rep(1, N)
## length of observation history from first capture
len <- T - first + 1
## survival probability
phi <- 0.7
## detection probability
pVec <- rep(c(0.2, 0.8), each=N/2)

## simulate z (alive / dead status),
## and y (encounter histories)
set.seed(0)
z <- matrix(NA, nrow=N, ncol=T)
y <- matrix(NA, nrow=N, ncol=T)
for(i in 1:N) {
    z[i, first[i]] <- y[i, first[i]] <- 1
    for(t in (first[i]+1):T) {
        z[i,t] <- rbinom(1, 1, phi*z[i,t-1])
        y[i,t] <- rbinom(1, 1, pVec[i]*z[i,t])
    }
}

## Homogeneous Capture-Recapture Model
code <-  nimbleCode({
    phi ~ dunif(0, 1)
    p ~ dunif(0, 1)
    for(i in 1:N) {
        y[i,first[i]:T] ~ dCJS_ss(phi, p, len=len[i])
    }
})
constants <- list(N=N, T=T, first=first, len=len)
data <- list(y=y)
inits <- list(phi=0.5, p=0.5)
Rmodel <- nimbleModel(code, constants, data, inits)

## 2-Group Finite Mixture Capture-Recapture Model
code <- nimbleCode({
    phi ~ dunif(0, 1)
    for(k in 1:K)   p[k] ~ dunif(0, 1)
    one ~ dconstraint(p[1] <= p[2])
    for(i in 1:N) {
        g[i] ~ dcat(pi[1:K])
        y[i,first[i]:T] ~ dCJS_ss(phi, p[g[i]], len=len[i])
    }
})
K <- 2    ## fixed number of groups
constants <- list(N=N, T=T, first=first, len=len, K=K)
data <- list(y=y, one=rep(1,K-1))
inits <- list(phi=0.5, pi=rep(1/K,K), p=rep(0.5,K), g=rep(1,N))
Rmodel <- nimbleModel(code, constants, data, inits)

## 3-Group Finite Mixture Capture-Recapture Model
code <- nimbleCode({
    phi ~ dunif(0, 1)
    for(k in 1:K)   p[k] ~ dunif(0, 1)
    one[1] ~ dconstraint(p[1] <= p[2])
    one[2] ~ dconstraint(p[2] <= p[3])
    for(i in 1:N) {
        g[i] ~ dcat(pi[1:K])
        y[i,first[i]:T] ~ dCJS_ss(phi, p[g[i]], len=len[i])
    }
})
K <- 3    ## fixed number of groups
constants <- list(N=N, T=T, first=first, len=len, K=K)
data <- list(y=y, one=rep(1,K-1))
inits <- list(phi=0.5, pi=rep(1/K,K), p=rep(0.5,K), g=rep(1,N))
Rmodel <- nimbleModel(code, constants, data, inits)

## Non-Parametric Capture-Recapture Model
code <- nimbleCode({
    phi ~ dunif(0, 1)
    alpha ~ dgamma(1, 1)
    xi[1:N] ~ dCRP(conc=alpha, size=N)
    for(i in 1:M)   p[i] ~ dunif(0, 1)
    for(i in 1:N) {
        y[i,first[i]:T] ~ dCJS_ss(phi, p[xi[i]], len=len[i])
    }
})
M <- 100   ## maximum number of subgroups
constants <- list(N=N, T=T, first=first, len=len, M=M)
data <- list(y=y)
inits <- list(phi=0.5, alpha=1, xi=rep(1,N), p=rep(0.5,M))
Rmodel <- nimbleModel(code, constants, data, inits)


##################################################
################ Occupancy Models ################
##################################################

## number of sites
N <- 4000
## number of observation periods
T <- 6
## probability of occupancy
pOcc <- 0.7
## detection probability
pVec <- rep(c(0.2, 0.8), each=N/2)

## simulate z (occupied status),
## and y (encounter histories)
set.seed(0)
z <- rep(NA, N)
y <- matrix(NA, nrow=N, ncol=T)
for(i in 1:N) {
    z[i] <- rbinom(1, size=1, prob=pOcc)
    y[i, 1:T] <- rbinom(T, size=1, prob=z[i]*pVec[i])
}

## Homogeneous Occupancy Model
code <- nimbleCode({
    pOcc ~ dunif(0, 1)
    p ~ dunif(0, 1)
    for(i in 1:N) {
        y[i,1:T] ~ dOcc_s(pOcc, p, len=T)
    }
})
constants <- list(N=N, T=T)
data <- list(y=y)
inits <- list(pOcc=0.5, p=0.5)
Rmodel <- nimbleModel(code, constants, data, inits)

## 2-Group Finite Mixture Occupancy Model
code <- nimbleCode({
    pOcc ~ dunif(0, 1)
    for(k in 1:K)   p[k] ~ dunif(0, 1)
    one ~ dconstraint(p[1] <= p[2])
    for(i in 1:N) {
        g[i] ~ dcat(pi[1:K])
        y[i,1:T] ~ dOcc_s(pOcc, p[g[i]], len=T)
    }
})
K <- 2    ## fixed number of groups
constants <- list(N=N, T=T, K=K)
data <- list(y=y, one=rep(1,K-1))
inits <- list(pOcc=0.5, pi=rep(1/K,K), p=rep(0.5,K), g=rep(1,N))
Rmodel <- nimbleModel(code, constants, data, inits)

## 3-Group Finite Mixture Occupancy Model
code <- nimbleCode({
    pOcc ~ dunif(0, 1)
    for(k in 1:K)   p[k] ~ dunif(0, 1)
    one[1] ~ dconstraint(p[1] <= p[2])
    one[2] ~ dconstraint(p[2] <= p[3])
    for(i in 1:N) {
        g[i] ~ dcat(pi[1:K])
        y[i,1:T] ~ dOcc_s(pOcc, p[g[i]], len=T)
    }
})
K <- 3    ## fixed number of groups
constants <- list(N=N, T=T, K=K)
data <- list(y=y, one=rep(1,K-1))
inits <- list(pOcc=0.5, pi=rep(1/K,K), p=rep(0.5,K), g=rep(1,N))
Rmodel <- nimbleModel(code, constants, data, inits)

## Non-Parametric Occupancy Model
code <- nimbleCode({
    pOcc ~ dunif(0, 1)
    alpha ~ dgamma(1, 1)
    xi[1:N] ~ dCRP(conc=alpha, size=N)
    for(i in 1:M)   p[i] ~ dunif(0, 1)
    for(i in 1:N) {
        y[i,1:T] ~ dOcc_s(pOcc, p[xi[i]], len=T)
    }
})
M <- 100   ## maximum number of subgroups
constants <- list(N=N, T=T, M=M)
data <- list(y=y)
inits <- list(pOcc=0.5, alpha=1, xi=rep(1,N), p=rep(0.5,M))
Rmodel <- nimbleModel(code, constants, data, inits)


##################################################
############## Fit Model Using MCMC ##############
##################################################

## configure MCMC
conf <- configureMCMC(Rmodel)

## build MCMC
Rmcmc <- buildMCMC(conf)

## compile model and MCMC
Cmodel <- compileNimble(Rmodel)
Cmcmc <- compileNimble(Rmcmc, project=Rmodel)

set.seed(0)
samplesList <- runMCMC(Cmcmc, niter=10000, nchains=3)
\end{verbatim}
\end{footnotesize}
\end{singlespace}

\end{document}